\documentclass[12pt]{article}
\usepackage[dvips]{color}
\usepackage{epsfig}
\usepackage{graphicx}
\usepackage{amsmath}

\begin{document}
\begin{center}
\Large{\bf The Lagrangian of charged test particle in Horava-Lifshitz black hole and deformed phase space }\\
\small \vspace{1cm} {\bf Z. Nekouee
\footnote{z.nekouee@stu.umz.ac.ir}},\quad{\bf J. Sadeghi
\footnote{pouriya@ipm.ir}}\quad and\quad{\bf M. Shokri \footnote{mehdi.shokri@uniroma1.it}}\\
\vspace{0.5cm}$^{1}${\it Department of Mathematic, Faculty of Mathematic,\\
University of Mazandaran, P. O. Box 47416-95447, Babolsar, Iran}\\
\vspace{0.5cm}$^{2}${\it Department of Physics, Faculty of Basic Sciences,\\
University of Mazandaran, P. O. Box 47416-95447, Babolsar, Iran}\\
\vspace{0.5cm}$^{3}${\it Physics Department and INFN,\\
Universit`a di Roma “La Sapienza”, Ple. Aldo Moro 2, 00185, Rome, Italy}\\
\end{center}\vspace{1.2cm}
\date{}

\begin{abstract}
In this paper, we use deformation approach and obtain the corresponding Lagrangian of charged test particle. Also, we show the effect of NC parameters on the Lagrangian of test particle in HL background with charge and without charge. Also, we see in case of $\beta=\theta$ and without charge, the deformed and non-deformed Lagrangian will be  same.  Also in case of $\beta=\theta$ and with charge  will be the same but the charge or field need some scaling. Finally, We show that two theories in case of $\beta\neq\theta$ with charge is completely different. It means that in case of NC geometry in addition to have time components of field we have $r$ and $x_{i}$ components.
\end{abstract}
\newpage
\section{Introduction}
One of the most significant problems in modern physics is a combination between general relativity (GR) and quantum mechanics (QM) to explain the gravity in all scales [1, 2]. The problem is open to now, but many attempts have done to make quantum gravity (QG) by diverse ways. Some people believe that we can access to QG theory in the context of string theory that it leads to AdS/CFT correspondence [3-5]. On the other hand, one can explain QG by considering diverse quantum corrections for GR [6]. One of the most interesting models introduced by Horava which is named Horava-Lifshtiz (HL) gravity [7, 8]. This approach to gravity can be a good candidate for UV completion and renormalization of GR by adding higher order spatial derivatives without adding higher-order time derivatives. According to the HL theory, we are faced with some anisotropic scaling relation $x \rightarrow bx$ and $t \rightarrow b^{z}t$ where $b$ is a constant and $z$ is responsible to measure the degree of anisotropy of space and time coordinates. HL gravity is based on the basic assumption that the Lorentz
symmetry is fundamentally broken at high scales of energy and restores only in the low limit. In contrast, some people believe that for each finite energy scale the Lorentz symmetry is broken and the HL gravity admits Lorentz symmetry preserving preferred global time foliation of the space-time. This symmetry can be related to the standard Lorentz transformations by the frame dependent change of synchronization to the Einstein one [9]. Because of the importance of HL gravity, there are many works that are specified for cosmological scenarios particularly black hole solutions of HL gravity. Even though, the investigation of HL black holes seems be simple asymptotically, but it can be very difficult when we are interested to present a analytic solution [10]. We can imply to some examples about this issue. In ref. [11], a $4D$ topological black hole with the dynamical exponent $z = 2$ investigated which behaves such a Lifshitz black hole asymptotically. Another example was associated to an analytic black hole solution with $z = 2$ in $4D$ space-times which was a Lifshitz spacetime asymptotically [12]. Also, its numerical solutions were investigated in [13, 14]. Moreover, The analytic examples of
Lifshitz black holes in higher dimensions were reported in [15]. The $z = 3$ Lifshitz black hole [16] was derived in the context of new massive gravity in $3D$ space-times [17].\\
Also, the present work is based on non-commutative (NC) geometry as a very important subject which is engaged to physical problems [18, 19]. In principle, this geometry is established on the idea of strong connection between non-commutativity and field theory. String theory attempts to explain quantum gravity in addition to solve the unification problem and eventually, it leads to a NC spacetime. As an example of this kind of geometry, we can imply to the phase space of ordinary quantum mechanics [20]. The momenta of a system in the presence of magnetic field are non-commuting operators as well. As a consequence of applying the non-commutativity between spatial and time coordinates, we might encounter with some problems in unitarity and causality, hence we usually just consider spatial NC. Besides, so far
quantum theory on the NC space has been extensively studied, the main approach is based
on the Weyl-Moyal product, in that case, the usual product replace by $\star$ in the NC space.
Therefore, deformation of spacetime plays a great role in the study of physical systems on
the NC space. Moreover, the problem of quantum mechanics on NC spaces can be found in the context of deformed spacetime [21, 22]. The NC space from the approach of deformation for harmonic oscillators is reported in [23, 24].\\
Our main focus for the application of NC geometry in cosmology is associated to the HL black holes. For the first step, they considered the non-commutative corrections for massive gravity black holes [25]. In the present work, we take advantage from above information to deform the Hamiltonian of HL black holes with non-commutative geometry space. In that case, we have some deformation parameter which are non-commutative geometry parameter. All above information give us motivation to arrange paper as follow. In section $2$ we are going to review general form of HL black hole. In section $3$ we take some information from section $2$ and write the explicit form of Lagrangian. Such Lagrangian is explanation of charged test particle in corresponding black holes. In order to apply NC geometry to Hamiltonian of black hole, one can choose some change of variables. We also write the non-deformed Hamiltonian corresponding to charged test particle in HL black hole. Then, we give some review to NC geometry and apply it to non-deformed Hamiltonian. So, the deformed Hamiltonian help us to achieve the deformed Hamiltonian for charged test particle in HL black hole. Finally, we present some results and conclusion.
\section{Horava-Lifshitz black holes}
In order to study the HL black holes, the general form of action can be written as,
\begin{equation}
S_{t}=S_{g}+S_{m}
\end{equation}
where $S_{g}$ and $S_{m}$ are gravitational and matter sectors, respectively. So, according to Ref. [26], the form of action for the gravitational section in HL gravity can be found as
\begin{equation}
S_{g}=\frac{1}{2\kappa^{2}}\int dtdrd^{d}x\sqrt{-G}(K_{ab}K^{ab}-\lambda K^{2}+\beta(R-2\Lambda)+\frac{\alpha^{2}}{2}\frac{\nabla_{a}N\nabla^{a}N}{N^{2}}),
\end{equation}
where we can see that the action will be reduced to GR action if $\lambda=\beta=1$ and $\alpha=0$. In our analysis, we start with a general form of metric as,
\begin{equation}
ds^{2}=-N^{2}dt^{2}+g_{ij}(dx^{i}-N^{i}dt)(dx^{j}-N^{j}dt).
\end{equation}
where, $N$, $N^{i}$ and $g_{ij}$ are the lapse function, the shift vector and the metric of the space-like
hypersurface, respectively. Therefore, we have the following relations $\sqrt{-G} = \sqrt{g}N$,
 $K_{ab}=\frac{1}{2N}(\partial_{t}g_{ab}-\nabla_{a}N_{b}-\nabla_{b}N_{a})$ and $K=g^{ab}K_{ab}$.\\ 
For matter sector, we use Lorentz-violating electromagnetism field as [27],
\begin{eqnarray}
S_{m}=-\frac{1}{2\kappa^{2}}\int dtdrd^{d}x\sqrt{g}N(\frac{2}{N^{2}}g^{ij}(F_{0i}-F_{ki}N^{k})(F_{0j}-F_{lj}N^{l})-F_{ij}F^{ij}&\!-&\!\nonumber\\\beta_{0}
-\beta_{1}a_{i}B^{i}-\beta_{2}B_{i}B^{i})
\end{eqnarray}
where,
\begin{equation}
F_{\mu\nu}=\partial_{\mu}A_{\nu}-\partial_{\nu}A_{\mu},\quad B^{i}=\frac{1}{2}\frac{\Gamma^{ijk}}{\sqrt{g}}F_{jk}
\end{equation}
with $\Gamma^{ijk}$ the Levi-Civita symbol.\\
Now, we are going to obtain the solution of Eqs. (2) and (4) first, we study the action of (4). In that case, we consider the electromagnetic field with the only non-vanishing component $A_{t}(r)$ and also
$\beta_{\mu}=0(\mu=0, 1, 2)$. Subsequently, the Maxwell equation reads as,
\begin{equation}
\partial_{r}(\sqrt{g}NF^{rt})=0,
\end{equation}
which gives us the following solution,
\begin{equation}
F^{rt}=\frac{Q_{e}}{\sqrt{g}N}
\end{equation}
with the integral constant $Q_{e}$, which can be interpreted as the charge of the Lifshitz black
hole. By using analytic process in Ref. [28], we set the metric
components in (3) as
\begin{equation}
N=e^{2f(r)},\quad g_{rr}=\frac{1}{e^{2h(r)}},\quad g_{ii}=e^{2l(r)},\quad N_{a}=0.
\end{equation}
By substitution of the above components in the total action (1), one can obtain
\begin{equation}
f(r)=z\ln r+\frac{1}{2}\ln\xi(r),\quad h=\ln r+\frac{1}{2}\ln\xi(r),\quad l=\ln r.
\end{equation}
then the metric (3) takes the following form,
\begin{equation}
ds^{2}=-r^{2z}\xi(r)dt^{2}+\frac{dr^{2}}{r^{2}\xi(r)}+r^{2}dx_{i}^{2}.
\end{equation}
and the form of $\xi(r)$ in the above equation will be following
\begin{equation}
\xi(r)=1-\frac{M}{r^{d+z}}+\frac{2zQ_{e}^{2}}{d\beta(d-z)}\frac{1}{r^{2d}}.
\end{equation}
where $M$ and $Q_{e}$ are integral constants which can be understood as the mass and charge
of the Lifshitz black hole.\\
By combination of Eqs. (5) and (7) for the time component of $A$, we will have
\begin{equation}
A_{t}(r)=\frac{Q_{e}}{d-z}(1-(\frac{r_{0}}{r})^{d-z}).
\end{equation}
\section{The deformation of phase space}
Firstly, we consider the general form of Lagrangian for the charged test
particle,
\begin{equation}
\mathcal{L}=\frac{1}{2}(g_{\mu\nu}\frac{dx^{\mu}}{d\tau}\frac{dx^{\nu}}{d\tau})+eA_{\mu}\frac{dx^{\mu}}{d\tau}.
\end{equation}
The conservation of angular momentum confines the particle motion to plane. Also we conveniently take the equation plane with $\theta=\frac{\pi}{2}$, $d\theta=0$ and $\sin\theta=1$. As a result of this fact and by using the metric (10) in the above equation, we can obtain the Lagrangian for HL black holes as
\begin{equation}
\mathcal{L}=\frac{1}{2}(-r^{2z}\xi(r)\dot{t}^{2}+\frac{\dot{r}^{2}}{r^{2}\xi(r)}+r^{2}\dot{x_{i}}^{2})+eA_{t}(r)\dot{t}.
\end{equation}
Now, we can apply the canonical relations in order to obtain the corresponding Hamiltonian by rewriting the Hamiltonian in terms of new variables. We note that the new form of Hamiltonian for the Lagrangian of HL black holes will be similar to the form of simple harmonic oscillator. We define new variables as,
\begin{equation}
\begin{array}{ccc}
x_{1}=r^{z}\sqrt{\xi(r)}\cosh t, & x_{2}=r\sinh x_{i}, & x_{3}=\frac{1}{r\sqrt{\xi(r)}}\sinh r, \\
y_{1}=r^{z}\sqrt{\xi(r)}\sinh t, & y_{2}=r\cosh x_{i}, & y_{3}=\frac{1}{r\sqrt{\xi(r)}}\cosh r,
\end{array}
\end{equation}
and
\begin{equation}
\begin{array}{cc}
x_{4}+y_{4}=\sqrt{2}r+r^{z}\sqrt{\xi(r)}, & x_{5}+y_{5}=\frac{1}{r\sqrt{\xi(r)}}+r,\\
x_{4}-y_{4}=\sqrt{2}r-r^{z}\sqrt{\xi(r)}, & x_{5}-y_{5}=\frac{1}{r\sqrt{\xi(r)}}-r,
\end{array}
\end{equation}
\begin{equation}
\begin{array}{cc}
x_{6}=\cosh\sqrt{eA_{t}(r)\dot{t}}, & x_{7}+y_{7}=\sqrt{eA_{t}(r)\dot{t}}+1,\\
y_{6}=\sinh\sqrt{eA_{t}(r)\dot{t}}, & x_{7}-y_{7}=\sqrt{eA_{t}(r)\dot{t}}-1.
\end{array}
\end{equation}
By using the above variables in Eq. (14), one can obtain the Hamiltonian as,
\begin{equation}
H=\frac{1}{2}\sum\limits_{i=1}^7 ((P_{x_{i}}^{2}-P_{y_{i}}^{2})+\omega_{i}^{2}(x_{i}^{2}-y_{i}^{2}))
\end{equation}
where
\begin{equation}
\begin{array}{c}
P_{x_{i}}=\frac{\partial\mathcal{L}}{\partial\dot{x_{i}}}=\dot{x_{i}},\\
P_{y_{i}}=\frac{\partial\mathcal{L}}{\partial\dot{y_{i}}}=-\dot{y_{i}}
\end{array}
\end{equation}
and
\begin{eqnarray}
\omega_{i}^{2}=-1,\quad i=1, 2, ..., 5 & \omega_{j}^{2}=0,\quad j=6, 7.
\end{eqnarray}
It is obvious that the Hamiltonian Eq. (18) has an oscillator form that
is useful for gravitational theories.\\
As we know, in commutative case we have usual Poisson brackets which are given by,
\begin{equation}
\{x_{i},x_{j}\}=0,\quad\{P_{x_{i}},P_{x_{j}}\}=0,\quad\{x_{i},P_{x_{j}}\}=\delta_{ij},
\end{equation}
where $x_{i}(i=1, 2)$ and $P_{x_{i}}(i=1, 2)$.  To compare NC and
commutative phase space
we need to explain some approaches of non-commutative. In such
approach, quantum effects can be dissolved by the Moyal brackets
$\{f,g\}_{\alpha}=f\star_{\alpha}g-g\star_{\alpha}f$ which is based
on the Moyal product as,
\begin{equation}
(f\star_{\alpha}g)(x)=exp[\frac{1}{2}\alpha^{ab}\partial_{a}^{(1)}\partial_{b}^{(2)}]f(x_{1})g(x_{2})|_{x_{1}=x_{2}=x}.
\end{equation}
After corresponding calculations, we find some anti commutative relations which are given by
\begin{equation}
\{x_{i},x_{j}\}_{\alpha}=\theta_{ij},\quad\{x_{i},P_{j}\}_{\alpha}=\delta_{ij}+\sigma_{ij},\quad\{P_{i},P_{j}\}=\beta_{ij}.
\end{equation}
Now, we focus on the NC geometry and apply it for new Hamiltonian of charged test particle in HL black hole. In that case, we use transformations on the classical phase space variables. So, we have following
\begin{equation}
\hat{x_{i}}=x_{i}+\frac{\theta}{2}P_{y_{i}},\quad \hat{y_{i}}=y_{i}-\frac{\theta}{2}P_{x_{i}},\quad \hat{P_{x_{i}}}=P_{x_{i}}-\frac{\beta}{2}y_{i},
\quad \hat{P_{y_{i}}}=P_{y_{i}}+\frac{\beta}{2}x_{i}.\\
\end{equation}
The new anti commutative relation will be as
\begin{equation}
\{\hat{y},\hat{x}\}=\theta,\quad\{\hat{x},\hat{P_{x}}\}=\{\hat{y},\hat{P_{y}}\}=1+\sigma,\quad\{\hat{P_{y}},\hat{P_{x}}\}=\beta,
\end{equation}
where $\sigma=\frac{\beta\theta}{2}$. In order to construct the
deformed Hamiltonian of charged test particle in HL black hole, we take the Hamiltonian
from Eq. (18) with new variables in Eq. (24). Hence, the form of
Hamiltonian in the deformed analysis is found as,
\begin{equation}
\hat{H}=\frac{1}{2}\sum_{i=1}^{7}((P_{x_{i}}^{2}-P_{y_{i}}^{2})-\gamma_{i}^{2}(y_{i}P_{x_{i}}+x_{i}P_{y_{i}})+
{\tilde{\omega_{i}}^{2}}(x_{i}^{2}-y_{i}^{2}))
\end{equation}
where
\begin{equation}
{\tilde{\omega_{i}}}^{2}=\frac{{\omega_{i}}^{2}-\frac{\beta^{2}}{4}}{1-{\omega_{i}}^{2}
\frac{\theta^{2}}{4}},\quad
{\gamma_{i}}^{2}=\frac{\beta-{\omega_{i}}^{2}\theta}{1-{\omega_{i}}^{2}\frac{\theta^{2}}{4}}.
\end{equation}
The above deformed Hamiltonian lead us to have a new Lagrangian which is deformed form
of original Lagrangian of charged test particle HL black hole. Now, for the corresponding model we arrange the
deformed Lagrangian as
\begin{equation}
\hat{\mathcal{L}}=\frac{1}{2}\sum_{i=1}^{7}((P_{x_{i}}^{2}-P_{y_{i}}^{2})-{\hat{\gamma_{i}}^{2}}(y_{i}P_{x_{i}}+x_{i}P_{y_{i}})-
{\hat{\omega_{i}}^{2}}(x_{i}^{2}-y_{i}^{2}))
\end{equation}
where
\begin{equation}
{\hat{\omega_{i}}}^{2}=\frac{{\omega_{i}}^{2}+\frac{\beta^{2}}{4}}{1+{\omega_{i}}^{2}
\frac{\theta^{2}}{4}},\quad
{\hat{\gamma_{i}}}^{2}=\frac{\beta+{\omega_{i}}^{2}\theta}{1+{\omega_{i}}^{2}\frac{\theta^{2}}{4}}.
\end{equation}
We use the Eq. (19), then the deformed Lagrangian can be written as following
\begin{eqnarray}
\hat{\mathcal{L}}=\frac{1}{2}[-r^{2z}\xi(r)\dot{t}^{2}+\frac{\dot{r}^{2}}{r^{2}\xi(r)}+r^{2}\dot{x_{i}}^{2}+
(\frac{(\beta-\theta)}{1-\frac{\theta^{2}}{4}}r^{2z}\xi(r)+\frac{\beta^{2}}{4}eA_{t}(r))\dot{t}-\frac{(\beta-\theta)}{1-\frac{\theta^{2}}{4}}(r^{2}\dot{x_{i}}&\!+&\!\nonumber\\
(\frac{1}{r^{2}\xi(r)}+\sqrt{2}(1-z)r^{z}\sqrt{\xi(r)}-\frac{\sqrt{2}r^{z+1}\acute{\xi}(r)}{2\sqrt{\xi(r)}}
-\frac{2}{r\sqrt{\xi(r)}}-\frac{\acute{\xi}(r)}{2\xi(r)\sqrt{\xi(r)}})\dot{r})].
\end{eqnarray}
Such equation is deformation form of Eq. (14) and the parameters $\beta$ and $\theta$ play important role in corresponding background. If we compare Eq. (30) with Eq. (14), we will arrive to the following results: For the first case, if we take $\beta=\theta$ in Eq. (30), the deformation will effect on the corresponding charge. In that case, the charge will be rescaled by $e\rightarrow\hat{e}=\frac{\beta^{2}}{4}e$ and the last term of Lagrangian will be $\frac{\beta^{2}}{4}eA_{t}(r)$ or the field may be changed by $A_{t}(r)\rightarrow \hat{A}_{t}(r)=\frac{\beta^{2}}{4}A_{t}(r)$. In the second case, if we assume $\beta=\theta$ the deformed and non-deformed Lagrangian without charge will be the same. The NC parameters have no effects to the corresponding Lagrangian. Finally, one can say if we take $\beta\neq\theta$ and the charge is non-zero, we will have completely different results. In that case, in addition to have $\hat{A}_{t}$ components of field we have $\hat{A}_{x_{i}}$ and $\hat{A}_{r}$ fields. So, the components of electromagnetic field will be as
\begin{equation}
\begin{array}{c}
\hat{A}_{t}=\frac{1}{2}(\frac{(\beta-\theta)}{1-\frac{\theta^{2}}{4}}r^{2z}\xi(r)+\frac{\beta^{2}}{4}eA_{t}(r)),\\
\hat{A}_{r}=-\frac{1}{2}\frac{(\beta-\theta)}{1-\frac{\theta^{2}}{4}}\sqrt{2}(1-z)r^{z}\sqrt{\xi(r)}-\frac{\sqrt{2}r^{z+1}\acute{\xi}(r)}{2\sqrt{\xi(r)}}
-\frac{2}{r\sqrt{\xi(r)}}-\frac{\acute{\xi}(r)}{2\xi(r)\sqrt{\xi(r)}},\\
\hat{A}_{x_{i}}=-\frac{1}{2}\frac{(\beta-\theta)}{1-\frac{\theta^{2}}{4}}r^{2}
\end{array}
\end{equation}
So, generally the NC geometry as Landau model account some electromagnetic field to the corresponding theory.
\section{Conclusion}
In this paper, we applied deformation analysis by NC geometry in order to obtain the corresponding Lagrangian of charged test particle. Also, we presented the effect of NC parameters on the mentioned Lagrangian in HL black holes with charge and without charge. We showed that in case of $\beta=\theta$ and without charge, the deformed and non-deformed Lagrangian were similar.  Also, in case of $\beta=\theta$ and with charge, it was the same but the charge or field need to some scaling. Finally, we showed that two scenarios in case of $\beta\neq\theta$ with charge are completely different. It means that in case of NC geometry in addition to have time components of field we have $r$ and $x_{i}$ components. Such results in future help us to investigate the relation between Finsler and non-commutative geometry for the test charged particle in HL black hole.

\end{document}